\documentclass[pra,twocolumn,superscriptaddress,showpacs,preprintnumbers,amsmath,amssymb]{revtex4}
\usepackage{}

\usepackage{bm}
\usepackage{bbm}
\usepackage{amssymb}
\usepackage{amsfonts}
\usepackage{epsfig,graphicx}
\usepackage{amstext}
\usepackage{amsmath}
\usepackage{graphicx}
\usepackage{times}
\usepackage{txfonts}
\usepackage{dcolumn}
\usepackage{color}
\usepackage{dsfont}

\newcommand{\iden}{\mathds{1}}
\newcommand{\tr}{\mathrm{tr}}

\begin{document}

\title{Finite-size scaling of coherence and steered coherence in the Lipkin-Meshkov-Glick model}

\author{Ming-Liang Hu}
\email{mingliang0301@163.com}
\affiliation{School of Science, Xi'an University of Posts and Telecommunications, Xi'an 710121, China}
\affiliation{Institute of Physics, Chinese Academy of Sciences, Beijing 100190, China}

\author{Fan Fang}
\affiliation{School of Electronic Engineering, Xi'an University of Posts and Telecommunications, Xi'an 710121, China}

\author{Heng Fan}
\email{hfan@iphy.ac.cn}
\affiliation{Institute of Physics, Chinese Academy of Sciences, Beijing 100190, China}
\affiliation{CAS Center for Excellence in Topological Quantum Computation, University of Chinese Academy of Sciences, Beijing 100190, China}
\affiliation{Songshan Lake Materials Laboratory, Dongguan 523808, China}

\begin{abstract}
Quantum coherence reflects the origin of quantumness and might be
capable of extracting the subtle nature of a system. We investigate
the ground-state coherence and steered coherence in the
Lipkin-Meshkov-Glick model and show that they detect faithfully the
quantum phase transitions of this model. Moreover, we carry out
scaling analysis on the coherence and steered coherence by means of
the continuous unitary transformation method and it is found that
the scaling exponents are uniquely determined by the phase region
of this model. These results may provide useful insights into the
mechanism underlying quantum criticality in many-body systems.
\end{abstract}


\maketitle

\section{Introduction} \label{sec:1}
Quantum phase transition (QPT) is a purely quantum phenomenon which
originates from quantum fluctuations \cite{QPT}. Besides the
traditional method, it has also been widely studied by virtue of
concepts borrowed from quantum information theory in the past two
decades. To be specific, this is mainly carried out based on singular
behaviors of the variety of quantum correlations. For example, the
ground-state entanglement of two carefully chosen sublattices or two
neighboring spins spotlight successfully the QPTs of certain spin
models \cite{nature,Osborne,Gusj1,Gusj2,Amico}. Additionally,
the quantum discord of two spins can also cleanly signal the QPTs in
some one-dimensional spin chain models \cite{Vedral,RPP,Werlang,Hu};
in particular, it signals the QPTs of the Heisenberg \textit{XXZ}
model even at finite temperatures \cite{Werlang}.

As the origin of quantumness and the fundamental property of quantum
states not available in classical physics, quantum coherence remains
one of the research focuses of the quantum theory since the beginning
of the 20th century and it may underlie the singularity of quantum
critical behaviors of various many-body systems. In particular,
since the formulation of the resource theory of coherence
\cite{coher,Plenio,Hu}, the study of QPTs from the perspective
of the ground-state coherence has been given more attention due to
its fundamentality. In fact, its feasibility has been validated
through studying those one-dimensional spin models, including the
spin-1/2 Ising and \textit{XX} models \cite{chenj}, the \textit{XY}
model \cite{Karpat,Qin,Leisg,Liyc}, the general \textit{XYZ} model
\cite{Ywl}, as well as the spin-1 \textit{XXZ} model \cite{spin1}.

Starting from the coherence measures, one can not only interpret
those already known quantum correlations \cite{coen1,coen2,coen3,
coqd1,coqd2} but also introduce other quantifiers of quantumness
\cite{asc1,asc2,asc3,Hux,msc}. One of these coherence-based
quantifiers is the steered coherence for a bipartite state
$\rho_{AB}$, including the average steered coherence (ASC) with
respect to the mutually unbiased bases \cite{asc1,asc2,asc3} and the
maximal steered coherence (MSC) in the eigenbasis of $\rho_B=\tr_A
\rho_{AB}$ \cite{msc}. The feasibility of the ASC in signaling QPTs
has been validated for certain spin systems \cite{pra032305,pssb},
and compared with entanglement and quantum discord, it has the
benefit of being long ranged, thus releases the strict restriction
on the choice of special spins for probing QPTs.

While the above works are mainly on the one-dimensional systems,
the coherence in the high-dimensional systems might exhibit richer
phenomena due to their high coordinate number. One of these models
is the Lipkin-Meshkov-Glick (LMG) model. It was first introduced
in nuclear physics \cite{LMG1,LMG2,LMG3} and more recent studies showed
that it can be simulated in trapped ions \cite{lmgsm1,lmgsm2},
nitrogen-vacancy center ensembles \cite{lmgsm3}, and
superconducting qubits \cite{lmgsm4}. It has also found
applications in quantum information processing \cite{spsq1,spsq2}.
For these reasons, we investigate coherence and steered coherence
in the LMG model with emphasis on their capability to characterize
the quantum criticality. We consider the following three cases:
(\romannumeral+1) the thermodynamic limit case by means of the
semiclassical approach, (\romannumeral+2) the isotropic case which is
exactly solvable with arbitrary system size $N$, and (\romannumeral+3)
the anisotropic case by utilizing the continuous unitary
transformation (CUT) technique \cite{CUT1,CUT2}. The results show
that both the coherence and steered coherence detect QPT of the LMG
model faithfully, and their dependence on the system size is found
to be scaled with different exponents in different phases. These
observations show a robust pathway to explore quantum criticality in
many-body systems by the coherence based indicators.

The structure of this paper is as follows. In Sec. \ref{sec:2}, we
recall some preliminaries on quantifying coherence and steered
coherence, then in Sec. \ref{sec:3}, we present their solutions for
the LMG model. Sec. \ref{sec:4} is devoted to analyzing their
behaviors and their scaling exponents in different phase regions.
Finally, we summarize the main results in Sec. \ref{sec:5}.

\section{Preliminaries} \label{sec:2}
In this section we recall briefly two measures of coherence defined
within the resource theoretic framework and the related
quantification of steered coherence. For a state described by the
density operator $\rho$, we consider its $l_1$ norm of coherence and
relative entropy of coherence given by \cite{coher}
\begin{equation}\label{eq2-1}
 C_{l_1}^{\{|i\rangle\}}(\rho)=\sum_{i\neq j}|\rho_{ij}|,~
 C_{r}^{\{|i\rangle\}}(\rho)=S(\rho_{\mathrm{diag}})-S(\rho),
\end{equation}
where the subscripts $l_1$ and $r$ indicate that the metrics
are the $l_1$ norm and relative entropy, respectively, while the
superscript $\{|i\rangle\}$ indicates the reference basis in which
the coherence is defined. Moreover, $\rho_{ij}$ denotes the
elements of $\rho$ in the reference basis $\{|i\rangle\}$,
$\rho_{\mathrm{diag}}$ is an operator obtained from $\rho$ by
replacing all its off-diagonal elements with zero, $S(\rho)=
-\tr(\rho\log_2\rho)$ is the von Neumann entropy of $\rho$, and
likewise for $S(\rho_{\mathrm{diag}})$.

Starting from the above coherence measures, one can then consider
the steered coherence for a two-qubit state $\rho_{AB}$, with qubit
\textit{A} (\textit{B}) holding by Alice (Bob). There are two
frameworks for such a problem. Within the first framework, it was
formulated by considering the three mutually unbiased observables
$\{\sigma_{x,y,z}\}$ (i.e., the Pauli operators). Specifically,
Alice first measures $\sigma_i$ on qubit $A$ and then informs Bob
of her choice $\sigma_i$ and outcome $a$, based on which Bob
measures the average coherence of the collapsed states of qubit $B$
with respect to the eigenbases of the two $\sigma_j\neq\sigma_i$
(as two different spin directions cannot be measured
simultaneously in experiments, this could be achieved by Bob's
randomly chosen eigenbases of the two $\sigma_j\neq\sigma_i$ with
equal probability for every round of Alice's  measurements). After
many rounds of Alice's local measurements of $\{\sigma_{x,y,z}\}$
with equal probability and the classical communication between Alice
and Bob, the ASC of qubit \textit{B} will be given by \cite{asc1}
\begin{equation}\label{eq2-2}
 C_\alpha^{\mathrm{asc}}(\rho_{AB})= \frac{1}{2}\sum_{i\neq j,a} p_{\Pi_i^a}
                                     C_\alpha^{\sigma_{j}}(\rho_{B|\Pi_i^a}),
\end{equation}
where $p_{\Pi_i^a}= \tr(\Pi_i^a\rho_{AB})$ defines the probability
of Alice's outcome $a$ when she measures $\sigma_i$, $\rho_{B|\Pi_i^a}
= \tr_A(\Pi_i^a\rho_{AB})/p_{\Pi_i^a}$ is the collapsed state of $B$,
$\Pi_i^\pm=(\iden\pm\sigma_i)/2$ is the measurement operator ($\iden$
is the $2\times 2$ identity operator), and $\alpha=l_1$ or $r$.
When $C_\alpha^{\mathrm{asc}}(\rho_{AB})$ is larger than a
threshold (i.e., the tight upper bound of the sum of the single-qubit
coherences in the three mutually unbiased bases), it is said that
there is a nonlocal advantage of quantum coherence (NAQC). In Ref.
\cite{asc1}, the authors showed that the separable $\rho_{AB}$ can
never achieve the NAQC, that is, the states achieving the NAQC form a
subset of the entangled states. This observation also holds for the
$(d\times d)$-dimensional states \cite{asc2}. In this sense, one may
view what the NAQC captures as a kind of bipartite quantum correlation
stronger than entanglement.

The second framework refers to the steered coherence on \textit{B}
with respect to the eigenbasis $\mathcal{B}=\{|\xi_i\rangle\}$ of
$\rho_B=\tr_A \rho_{AB}$. To be explicit, Alice first performs the
positive-operator-valued measurements (POVM) $M$ on her qubit
\textit{A}; Bob then measures the coherence of $\rho_{B|M}=
\tr_A(M\rho_{AB})/p_M$ in the basis $\mathcal{B}$, where $p_M=
\tr(M\rho_{AB})$. The MSC on qubit \textit{B} can be written as
\cite{msc}
\begin{equation}\label{eq2-3}
 C_\alpha^{\mathrm{msc}}(\rho_{AB})= \inf_{\mathcal{B}}\left\{\max_{M\in\mathrm{POVM}}
                                     C_\alpha^{\mathcal{B}}(\rho_{B|M})\right\},
\end{equation}
where the infimum over $\mathcal{B}$ is necessary only when $\rho_B$
is degenerate. $C_\alpha^{\mathrm{msc}}(\rho_{AB})$ is also
intimately related to quantum correlations, e.g., it is maximal for
any pure entangled state with full Schmidt rank and vanishes for
quantum-classical states \cite{msc}.

The reasons we consider the coherence based indicators are as
follows. First, quantum coherence underlies entanglement and discord,
which have brought new insights to QPTs, thus it is quite likely that
the coherence based indicators would signal the QPTs. Second, they
are analytically solvable, whereas the calculation of discord is
rather more involved \cite{Vedral}. Finally, the critical points
captured by discord do not always correspond to QPTs as they may
stem from the sudden change of the optimal measurement basis in its
definition \cite{Vedral}; thus it is necessary to seek complementary
indicators of QPTs.

\section{Solution of coherence and steered coherence for the LMG model} \label{sec:3}
Spin system is a natural playground for revealing quantum
characteristics and is also a candidate for building blocks
implementing quantum computation and information processing tasks.
The LMG model characterizes a system consisting of $N$ spin-1/2
particles that are mutually interacted with each other. The
Hamiltonian (in units of $\hbar$) reads \cite{CUT1,CUT2}
\begin{equation}\label{eq3-1}
\begin{aligned}
 \hat{H}&= -\frac{\lambda}{N}\sum_{i<j}(\sigma_{x}^i \sigma_{x}^j+\gamma\sigma_{y}^i \sigma_{y}^j)-h\sum_i \sigma_{z}^i \\
        &= -\frac{2\lambda}{N}(S_x^2+\gamma S_y^2)-2h S_z+\frac{\lambda}{2}(1+\gamma),
\end{aligned}
\end{equation}
where $\sigma_{\upsilon}^i$ ($\upsilon=x,y,z$) represents the Pauli
operators at site $i$, $\lambda$ characterizes the spin interaction
(with $\gamma$ being its anisotropy in the $xy$ plane), and the
prefactor $1/N$ is introduced for eliminating infinity of the free
energy per spin in the thermodynamic limit. Moreover, $h$ is the
transverse magnetic field and $S_\upsilon=\sum_i \sigma_{\upsilon}^i/2$.
In the following, we concentrate on the ferromagnetic case and we set
$\lambda= 1$ (i.e., $h$ will be in units of $\lambda$), $0\leqslant
\gamma \leqslant 1$, and $h\geqslant 0$ (the spectrum of $\hat{H}$ is
invariant under the substitution $h \rightarrow -h$).

For such a model, its ground state lies in the maximum spin sector
$S=N/2$ \cite{CUT1,CUT2}. In the basis $\{\mid\uparrow\uparrow\rangle,
\mid\uparrow\downarrow\rangle, \mid\downarrow\uparrow\rangle,
\mid\downarrow\downarrow\rangle\}$, the two-spin reduced density
matrix can be obtained as \cite{RDM}
\begin{equation}\label{eq3-2}
 \rho_{ij}=
  \left(\begin{array}{cccc}
    v_1   & 0   & 0   & u \\
    0     & y   & y   & 0 \\
    0     & y   & y   & 0 \\
    u     & 0   & 0   & v_2 \\
  \end{array}\right) 
\end{equation}
and its elements $v_{1,2}$, $y$, and $u$ are given by
\begin{equation}\label{eq3-3}
\begin{aligned}
 & v_{1,2}= \frac{N^2-2N+4\langle S_z^2\rangle\pm 4(N-1)\langle S_z\rangle}{4N(N-1)}, \\
 & y= \frac{N^2-4\langle S_z^2\rangle}{4N(N-1)},~
   u= \frac{\langle S_x^2\rangle- \langle S_y^2\rangle}{N(N-1)},
\end{aligned}
\end{equation}
where $v_1$ takes the plus sign and $v_2$ takes the minus sign
(likewise for subsequent similar notation). The expectation values
$\langle S_z\rangle$ and $\langle S_\upsilon^2\rangle$ ($\upsilon=x,y,z$)
could be obtained by virtue of different methods for the Hamiltonian
\eqref{eq3-1} with different system parameters, and for $0\leqslant
\gamma \leqslant 1$ one also has $u\geqslant 0$ \cite{CUT1}.

From Eq. \eqref{eq3-2} one can obtain directly that there is no
single-spin coherence in the standard basis $\{\mid\uparrow\rangle,
\mid\downarrow\rangle\}$ as $\rho_i= \tr_j\rho_{ij}$ is diagonal.
Of course, the resource theoretic measures of coherence are
basis dependent \cite{coher}, so the coherence of $\rho_i$ measured
with other bases can be nonzero. For example, the maximum $l_1$
norm and relative entropy of coherences attainable by optimizing the
reference basis are given by \cite{MC}
\begin{equation}\label{eq3-n4}
 C_{l_1}^{\max}(\rho_i)= |v_1-v_2|,~
 C_{r}^{\max}(\rho_i)= 1-H_2(v_1+y),
\end{equation}
where $H_2(\cdot)$ denotes the binary Shannon entropy function.
For the two-spin state $\rho_{ij}$, the coherence can be obtained as
\begin{equation}\label{eq3-4}
\begin{aligned}
 & C_{l_1}(\rho_{ij})= 2(y+u), \\
 & C_{r}(\rho_{ij})= 2y +\sum_{i=1}^2(\epsilon_i\log_2\epsilon_i-v_i\log_2 v_i),
\end{aligned}
\end{equation}
while the ASC and MSC can be obtained as (see Appendix A)
\begin{equation}\label{eq3-5}
\begin{aligned}
 & C_{l_1}^{\mathrm{asc}}(\rho_{ij})=\sum_{i=1}^2 \left(\frac{1}{2}x_i +|y-v_i|\right)+|y+u|+|y-u|, \\
 & C_{r}^{\mathrm{asc}}(\rho_{ij})= 2 + H_2(y+v_1) \\
   &~~~~~~~~~~~~~~~~~ -\sum_{i=1}^2 \left[H_2\left(\frac{1+x_i}{2}\right)+(y+v_i)H_2\left(\frac{y}{y+v_i}\right)\right],\\
 & C_{l_1}^{\mathrm{msc}}(\rho_{ij})= \frac{2(y+u)}{\sqrt{1-(v_1-v_2)^2}}, \\
 & C_{r}^{\mathrm{msc}}(\rho_{ij})= H_2(r_{11})-H_2\left(\frac{1+\sqrt{(1-2r_{11})^2+4|r_{12}|^2}}{2}\right),\\
\end{aligned}
\end{equation}
where the parameters $\epsilon_{1,2}$, $x_{1,2}$, $r_{11}$, and
$r_{12}$ are given by
\begin{equation}\label{eq3-6}
\begin{split}
 & \epsilon_{1,2}=\frac{1}{2}\left[v_1+v_2\pm\sqrt{4u^2+(v_1-v_2)^2}\right],\\
 & x_{1,2}=\sqrt{(v_1-v_2)^2+ 4(y\pm u)^2},~
   r_{12}= \frac{y+u}{\sqrt{1-(v_1-v_2)^2}},\\
 & r_{11}= \frac{v_1}{1+v_1-v_2}+\frac{y}{1-v_1+v_2}.
\end{split}
\end{equation}

Starting from the above analytical results, we explore in the
following the performances of the coherence based indicators of
the single- and two-spin reduced density matrices in signaling
the occurrence of QPT in the LMG model.

\section{Results} \label{sec:4}
In the past few years, the quantum correlations in the LMG model,
including the two-spin entanglement \cite{ent1,ent2,ent3,ent4},
entanglement entropy \cite{entro1,entro2,entro3}, multipartite
entanglement \cite{men1,men2,men3}, multipartite nonlocality
\cite{nonl}, and quantum discord \cite{qd1,qd2}, have been
studied in depth in the context of QPTs. Moreover, the fidelity
susceptibility \cite{fs1,fs2}, quantum Fisher information \cite{QFI},
and the Loschmidt echo and fidelity \cite{LE} have also been
widely studied.

As quantum coherence is a fundamental property of quantum states and
reflects the origin of quantum correlations \cite{coen1,coen2,coen3,
coqd1,coqd2}, it is natural to speculate that the quantum coherence
may also play a vital role in the understanding of the quantum
criticality in the high-dimensional systems. In particular, the
two-spin entanglement in the LMG model decays rapidly with the increase
of the system size $N$ for any $h$ and disappears in the large-$N$
limit \cite{CUT2}, while the calculation of the multipartite entanglement
and nonlocality are intractable \cite{men1,men2,men3,nonl}. Quantum
coherence and steered coherence for any two spins in the bulk, however,
are not so sensitive to the increasing system size and are analytically
solvable. Thereby, it is expected that they may discover useful
information of the many-body systems. Motivated by these observations,
we investigate in this section the coherence, ASC, and MSC in the LMG
model and carry out a scaling analysis of their dependence on the
system size $N$ in different phase regions.

\subsection{The thermodynamic limit} \label{sec:4a}
In the thermodynamic limit, a semiclassical approach gives the
mean-field wave function \cite{mfa1,mfa2}
\begin{equation}\label{eq4-1}
 |\psi(\theta,\phi)\rangle= \mathop{\otimes}\limits_{l=1}^N \big[\cos(\theta/2)e^{-\mathrm{i}\phi/2}\mid\uparrow\rangle_l
                            + \sin(\theta/2)e^{\mathrm{i}\phi/2}\mid\downarrow\rangle_l \big],
\end{equation}
with $\mid\uparrow\rangle_l$ and $\mid\downarrow\rangle_l$ the
eigenstates of $\sigma_{z}^l$ with eigenvalues $1$ and $-1$,
respectively. Then one has
\begin{equation}\label{eq4-2}
 \langle \hat{H}\rangle= -\frac{(N-1)}{2}\sin^2\theta (\cos^2\phi+\gamma\sin^2\phi)-hN\cos\theta,
\end{equation}
and the ground state can be obtained by minimizing
$\langle \hat{H}\rangle$ over $\theta$ and $\phi$. For such a model,
a second-order QPT occurs at the critical point $h=1$. In the region
of $h\geqslant 1$ (symmetric phase), the optimal angle $\theta_0=0$,
all the spins are fully polarized in the direction of external
magnetic field. In the region of $0\leqslant h<1$ (broken phase),
$\theta_0= \arccos h$, the ground state will be twofold degenerate
($\phi_0=0$ or $\pi$) for $\gamma\neq 1$ and infinitely degenerate
($\phi_0$ can be arbitrary angle) for $\gamma=1$.

We focus on $\gamma<1$ (the case of $\gamma=1$ is exactly solvable
for any $N$ and will be discussed in the next subsection). From Eq.
\eqref{eq4-1} one can obtain that when $h\geqslant 1$, $v_1= 1$,
$v_2= y= u= 0$. As a result, $C_\alpha(\rho_{ij}^\mathrm{T})=
C_\alpha^{\mathrm{msc}}(\rho_{ij}^\mathrm{T})=0$ and
$C_\alpha^{\mathrm{asc}}(\rho_{ij}^\mathrm{T})=2$ ($\alpha=l_1$
or $r$), where we have denoted by $\rho_{ij}^\mathrm{T}$ the
two-spin density operator in the thermodynamic limit. Hence when
the system is in the symmetric phase, the three quantumness measures
always remain constant, which is understandable as all the spins are
fully polarized in the magnetic field direction and the system's
ground state is $\mid\uparrow\uparrow\ldots\uparrow\rangle$.
Moreover, at first glance, it seems contradictory for the MSC to be
zero and the ASC to be finite in the symmetric phase. In fact,
there is no contradiction because they are defined in different bases:
One is in the eigenbasis of $\rho_j$ which is state dependent and
the other one is with respect to the three mutually unbiased bases.

When $0\leqslant h< 1$, $v_{1,2}=h_\pm^2/4$ and $y=u= (1-h^2)/4$,
where we have defined $h_\pm= 1\pm h$. Then, by making use of Eqs.
\eqref{eq3-4}, \eqref{eq3-5}, and \eqref{eq3-6}, one can obtain
\begin{equation}\label{eq4-3}
\begin{aligned}
 & C_{l_1}(\rho_{ij}^\mathrm{T})= \big[C_{l_1}^{\mathrm{msc}}(\rho_{ij}^\mathrm{T})\big]^2= 1-h^2,~
   C_{r}^{\mathrm{msc}}(\rho_{ij}^\mathrm{T})= H_2\left(\frac{h_{+}}{2}\right), \\
 & C_{r}(\rho_{ij}^\mathrm{T})= 1+\frac{1+h^2}{2}\log_2 (1+h^2)-\sum_{i=+,-}\frac{h_i^2\log_2 h_i}{2}, \\
 & C_{l_1}^{\mathrm{asc}}(\rho_{ij}^\mathrm{T})= \frac{1-h^2+3h+\sqrt{1+h^4-h^2}}{2}, \\
 & C_{r}^{\mathrm{asc}}(\rho_{ij}^\mathrm{T})= 2-H_2\left(\frac{h_+}{2}\right)-H_2\left(\frac{1+\sqrt{1+h^4-h^2}}{2}\right).
\end{aligned}
\end{equation}

\begin{figure}
\centering
\resizebox{0.40 \textwidth}{!}{%
\includegraphics{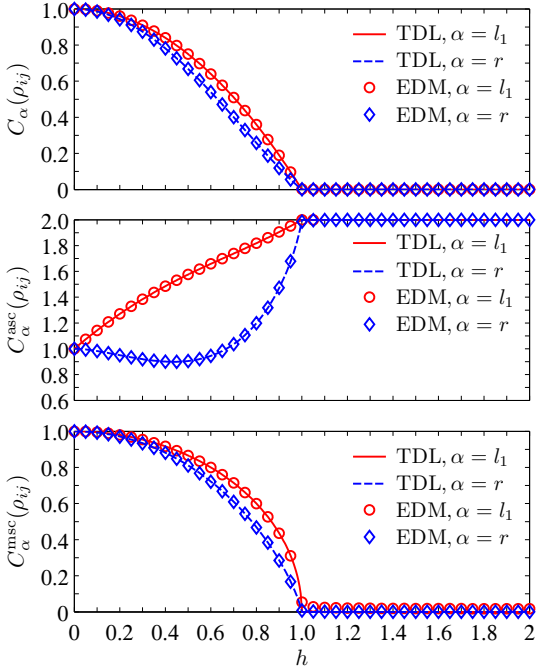}}
\caption{Plot of $C_{\alpha}(\rho_{ij})$, $C_{\alpha}^{\mathrm{asc}}(\rho_{ij})$,
and $C_{\alpha}^{\mathrm{msc}}(\rho_{ij})$ ($\alpha=l_1$ or $r$)
versus $h$, where the solid lines denote the thermodynamic limit (TDL)
results and the circles denote the exact diagonalization method (EDM)
results with $\gamma=0.5$ and $N=2^{12}$.} \label{fig:1}
\end{figure}

In Fig. \ref{fig:1} we show $h$ dependence of
$C_{\alpha}(\rho_{ij})$, $C_{\alpha}^{\mathrm{asc}}(\rho_{ij})$, and
$C_{\alpha}^{\mathrm{msc}}(\rho_{ij})$ for the thermodynamic limit
result and the finite-size result of the exact diagonalization
method. In the symmetric phase, as analyzed above, they always remain
constant. In the broken phase, however, both $C_{\alpha}(\rho_{ij})$
and $C_{\alpha}^{\mathrm{msc}}(\rho_{ij})$ ($\alpha=l_1$ or
$r$) decrease with the increase of $h$, whereas
$C_{l_1}^{\mathrm{asc}}(\rho_{ij})$ increases with the increase of
$h$, and $C_{r}^{\mathrm{asc}}(\rho_{ij})$ first decreases to a
minimum of about 0.8991 and then turns to be increased to 2. The
coherence, ASC, and MSC could therefore be used as reliable indicators
of QPT in this model. From Fig. \ref{fig:1} one can also note that the
finite-size results with $N=2^{12}$ are in good agreement with the
mean-field results which are adaptive for the thermodynamic limit case.
Similarly, for the single-spin state $\rho_i$, one has
$C_{\alpha}^{\max}(\rho_i^\mathrm{T})=1$ ($\alpha=l_1$ or $r$) for
$h\geqslant 1$, while $C_{l_1}^{\max}(\rho_i^\mathrm{T})=h$ and
$C_{r}^{\max}(\rho_i^\mathrm{T})= 1-H_2(h_{+}/2)$ for
$0\leqslant h< 1$. They also exhibit distinct behaviors in the two
different phases.

\subsection{The isotropic case} \label{sec:4b}
For the isotropic LMG model (i.e., $\gamma=1$), due to the global
symmetries described by $[\hat{H},S^2]=0$ and $[\hat{H},S_z]=0$,
one has $\langle\hat{H}\rangle = 2M^2/N-2hM-N/2$, with
$M= \langle S_z\rangle$. By minimizing $\langle \hat{H}\rangle$
over all $M=-N/2,\ldots,N/2$, one can obtain immediately the ground
state as $|S,M_0\rangle$, with
\begin{equation}\label{eq4-4}
 M_0=\left\{
  \begin{aligned}
   &N/2 &&\text{for}~h\geqslant 1, \\
   &I(hN/2) &&\text{for}~0\leqslant h<1,
  \end{aligned} \right.
\end{equation}
where $I(x)$ is the nearest integer (half-integer) from $x$ for even
(odd) $N$. When $h\geqslant 1$, $v_1=1$ and $v_2=y=u=0$, so one still
has $C_\alpha(\rho_{ij}) =C_\alpha^{\mathrm{msc}}(\rho_{ij})=0$ and
$C_\alpha^{\mathrm{asc}}(\rho_{ij})=2$ ($\alpha=l_1$ or $r$)
in the symmetric phase region.

When $0\leqslant h< 1$, by substituting $M_0$ into Eq. \eqref{eq3-3}
and using  $\langle S_x^2\rangle=\langle S_y^2\rangle$ for the ground
state $|S,M_0\rangle$, one can obtain
\begin{equation}\label{eq4-5}
\begin{aligned}
 & C_{l_1}(\rho_{ij})= C_{r}(\rho_{ij})= \frac{N_{+}N_{-}}{2N(N-1)},
   C_{l_1}^{\mathrm{msc}}(\rho_{ij})= \frac{\sqrt{N_{+}N_{-}}}{2N-2}, \\
 & C_{r}^{\mathrm{msc}}(\rho_{ij})= H_2\left(\frac{N_{+}-1}{2N-2}\right)-H_2\left(\frac{1}{2}+\frac{\sqrt{N^2+12M_0^2}}{4N-4}\right), \\
 & C_{l_1}^{\mathrm{asc}}(\rho_{ij})= x_0 +\frac{N_{+}|1-2M_0|+N_{-}(N+4M_0+1)}{2N(N-1)}, \\
 & C_{r}^{\mathrm{asc}}(\rho_{ij})= 2-2H_2\left(\frac{1+x }{2}\right) +H_2\left(\frac{N_{+}}{2N}\right) \\
 &~~~~~~~~~~~~~~~~~ - \frac{N_{+}}{2N}H_2\left(\frac{N_{-}}{2N-2}\right)-\frac{N_{-}}{2N}H_2\left(\frac{N_{+}}{2N-2}\right),
\end{aligned}
\end{equation}
where we have denoted by $N_\pm=N\pm 2M_0$ and $x_0=\big[N_{+}^2N_{-}^2
+16M_0^2(N-1)^2\big]^{1/2}/2N(N-1)$. From Eq. \eqref{eq4-5} one can
obtain that for large $N$, the coherence, ASC, and MSC show qualitatively
the same $h$ dependence as those shown in Fig. \ref{fig:1}.

\subsection{Finite-size scaling} \label{sec:4c}
For the general LMG model with a finite number of spins, the
expectation values $\langle S_z \rangle$ and
$\langle S_\upsilon^2 \rangle$ ($\upsilon=x,y,z$) can be obtained
approximately by means of the CUT method \cite{cutt1,cutt2,cutt3}.
Different from the Holstein-Primakoff transformation method based on
a first-order correction in the $1/N$ expansion of the spin operators,
the CUT method takes into account the higher-order corrections and thus
enables the extraction of the finite-size scaling exponents of
different observables \cite{CUT1,CUT2}. In the following, we first
figure out the scaling formulas of the ground-state coherence, ASC,
and MSC with the help of the CUT results of $\langle S_z\rangle$ and
$\langle S_\upsilon^2\rangle$ ($\upsilon = x,y,z$) and then confirm
them numerically using the exact diagonalization method with the
system size up to $N= 2^{16}$.

\begin{table}[!h]
\tabcolsep 0pt
\caption{Coefficients $a_\Phi$ ($\Phi=z,xx,yy,zz$) obtained numerically
         with $N=2^{16}$ and different $\gamma$.} \label{tab:1}
\vspace*{-12pt}
\begin{center}
\def\temptablewidth{0.48\textwidth}
{\rule{\temptablewidth}{1pt}}
\begin{tabular*}{\temptablewidth}{@{\extracolsep{\fill}}cclll}
       $a_\Phi$  &$\gamma=0$  &$\gamma=0.25$  &$\gamma=0.50$  &$\gamma=0.75$ \\   \hline
       $a_z$     &$-0.4599$   &$-0.4182$      &$-0.3659$      &$-0.2913$  \\
       $a_{xx}$  &0.9188      &0.8354         &0.7307         &0.5813  \\
       $a_{yy}$  &1.1144      &1.2257         &1.4017         &1.7621  \\
       $a_{zz}$  &$-0.9195$   &$-0.8362$      &$-0.7315$      &$-0.5824$ \\
\end{tabular*}
{\rule{\temptablewidth}{1pt}}
\end{center}
\end{table}

First of all, we consider scaling behaviors at the QPT point. The
expectation values of $\langle S_z\rangle$ and $\langle S_\upsilon^2\rangle$
($\upsilon=x,y,z$) obtained via the CUT technique can be found in Ref.
\cite{CUT1}. For $h=1$ with very large but finite $N$, by considering
the fact that there should be no singularity for any physical quantity
in a finite system, one can obtain from Ref. \cite{CUT1} that
\begin{equation}\label{eq4-6}
\begin{aligned}
 & \frac{2\langle S_z\rangle}{N} \sim 1+\frac{1}{N}+\frac{a_z}{N^{2/3}}, ~
   \frac{4\langle S_x^2\rangle}{N^2} \sim \frac{a_{xx}}{N^{2/3}}, \\
 & \frac{4\langle S_y^2\rangle}{N^2} \sim \frac{a_{yy}}{N^{4/3}}, ~
   \frac{4\langle S_z^2 \rangle}{N^2} \sim 1+\frac{2}{N}+\frac{a_{zz}}{N^{2/3}},
\end{aligned}
\end{equation}
where $a_\Phi$ ($\Phi=z,xx,yy,zz$) are the constants dependent on
$\gamma$. Although their values cannot be determined with this
scaling argument, we can obtain their approximate values by combining
the above equation with the numerically obtained $\langle S_z\rangle$
and $\langle S_\upsilon^2\rangle$ ($\upsilon=x,y,z$) with very large $N$;
see, e.g., the results listed in Table \ref{tab:1}, from which one can
note that $a_{zz}\simeq 2a_z$. Hence, by combining Eq. \eqref{eq4-6}
with Eq. \eqref{eq3-3} and neglecting the exponentially small terms
in $1/N$, one can obtain
\begin{equation}\label{eq4-7}
\begin{aligned}
 & y+u \sim \frac{a_{xx}}{2N^{2/3}},~        y-u \sim -\frac{1}{2N},~  \\
 & y+v_1 \sim 1+\frac{a_z}{2N^{2/3}},~       y-v_1 \sim -1-\frac{a_z +a_{zz}}{2N^{2/3}}, \\
 & y+v_2 \sim -\frac{a_z}{2N^{2/3}},~        y-v_2 \sim \frac{a_z -a_{zz}}{2N^{2/3}}, \\
 & v_1+v_2 \sim 1+\frac{a_{zz}}{2N^{2/3}},~  v_1-v_2 \sim 1+\frac{a_z}{N^{2/3}}.
\end{aligned}
\end{equation}

\begin{figure}
\centering
\resizebox{0.41 \textwidth}{!}{%
\includegraphics{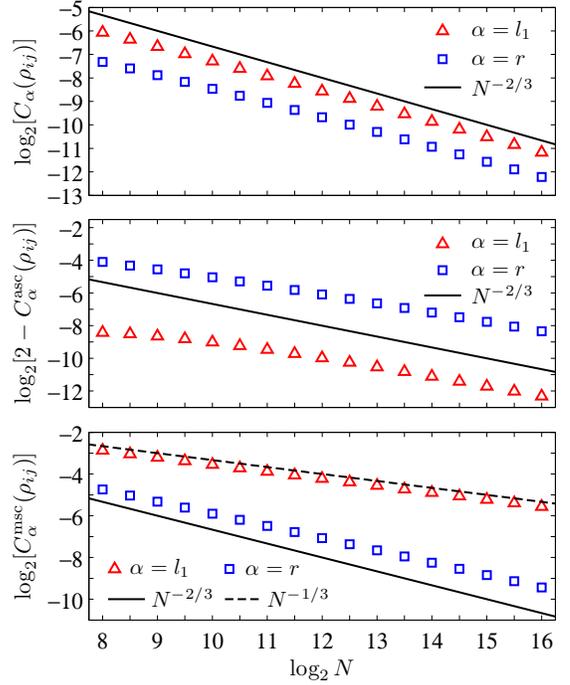}}
\caption{Scaling behaviors of $C_{\alpha}(\rho_{ij})$,
$C_{\alpha}^{\mathrm{asc}}(\rho_{ij})$, and
$C_{\alpha}^{\mathrm{msc}}(\rho_{ij})$ ($\alpha=l_1$ or $r$) at the
critical point $h=1$ with $\gamma=0.5$. The slopes of the solid and dashed
lines are $-2/3$ and $-1/3$, respectively.} \label{fig:2}
\end{figure}

When considering the $l_1$ norm and relative entropy of coherence for
the system at the critical point $h=1$, as $v_1-v_2\gg u$ for very
large $N$, one can obtain from Eq. \eqref{eq3-4} that
\begin{equation}\label{eq4-8}
 C_{l_1}(\rho_{ij}) \sim \frac{a_{xx}}{N^{2/3}},~
 C_{r}(\rho_{ij}) \sim -\frac{a_{zz}}{2N^{2/3}},
\end{equation}
thus the coherence of the two-spin state scales as
\begin{equation}\label{eq4-9}
 \log_2[C_{\alpha}(\rho_{ij})]\sim -\frac{2}{3}\log_2 N+ \mathrm{const},
\end{equation}
where $\alpha=l_1$ or $r$, and the constant depends on $\alpha$ and
$\gamma$ (likewise for those subsequent scaling formulas in
other phase regions). Such a scaling behavior could be verified by
diagonalizing the Hamiltonian $\hat{H}$ numerically in the spaces
spanned by $\{|S, M_1\rangle\}$ ($M_1= -N/2, -N/2+2, \ldots, N/2$)
and $\{|S,M_2\rangle\}$ ($M_2= -N/2+1, -N/2+3, \ldots, N/2-1$) in which
the ground state lies. For the system size up to $N= 2^{16}$, we
display in the topmost panel of Fig. \ref{fig:2} the numerical results,
from which one can note that $\log_2 [C_{\alpha}(\rho_{ij})]$
approaches the solid line with slope $-2/3$ with an increase in
$N$.

For the ASC, by using the approximation $(1+x)^{1/2}\simeq 1+x/2$ for
very small $x$, one can obtain
\begin{equation}\label{eq4-10}
\begin{aligned}
 & x_{1,2} \sim 1+\frac{a_z}{N^{2/3}},~
  \frac{y}{y+v_1}\sim -\frac{a_{zz}}{4N^{2/3}}, \\
 & \frac{y}{y+v_2}\sim \frac{a_{zz}}{2a_z}+\frac{1}{a_zN^{1/3}},
\end{aligned}
\end{equation}
and when the small $x$ is positive, one also has $H_2(x)\simeq x/\ln 2$.
These, together with Eqs. \eqref{eq3-5} and \eqref{eq4-7}, yield
\begin{equation}\label{eq4-11}
 C_{l_1}^{\mathrm{asc}}(\rho_{ij})\sim  2+\frac{4a_z+a_{xx}}{2N^{2/3}},~
 C_{r}^{\mathrm{asc}}(\rho_{ij})\sim  2 + \frac{2a_z +a_{zz}}{4N^{2/3}\ln 2},\\
\end{equation}
where we have used the facts that $a_{xx}>0$ and $a_{zz}\simeq 2a_z <0$ (Table \ref{tab:1}).
So the ASC has the finite-size scaling behavior 
\begin{equation}\label{eq4-12}
 \log_2 [2-C_{\alpha}^{\mathrm{asc}}(\rho_{ij})]\sim -\frac{2}{3}\log_2 N +\mathrm{const},
\end{equation}
where $\alpha=l_1$ or $r$. This scaling formula was
verified numerically by using the exact diagonalization method. As
is shown in the middle panel of Fig. \ref{fig:2}, the slope of
$\log_2[2-C_{\alpha}^{\mathrm{asc}}(\rho_{ij})]$ approaches $-2/3$
gradually with an increase in $N$.

For the MSC, from Eqs. \eqref{eq3-6} and \eqref{eq4-7} one can obtain
\begin{equation}\label{eq4-13}
 r_{11}= \frac{a_{zz}+2a_z}{4a_z}+\frac{a_{zz}+a_z}{4N^{2/3}},~
 r_{12}= \frac{a_{xx}}{2\sqrt{-2a_z}N^{1/3}},
\end{equation}
then by substituting these into Eq. \eqref{eq3-5} and using the fact
$a_{zz}\simeq 2a_z$, one can obtain
\begin{equation}\label{eq4-14}
 C_{l_1}^{\mathrm{msc}}(\rho_{ij})\sim \frac{a_{xx}}{\sqrt{-2a_z}N^{1/3}},~
 C_{r}^{\mathrm{msc}}(\rho_{ij})\sim -\frac{a_{xx}^2}{4a_{zz} N^{2/3} \ln 2},
\end{equation}
which indicate that the two forms of MSC scale as
\begin{equation}\label{eq4-15}
\begin{aligned}
 & \log_2 [C_{l_1}^{\mathrm{msc}}(\rho_{ij})] \sim -\frac{1}{3}\log_2 N+ \mathrm{const}, \\
 & \log_2 [C_{r}^{\mathrm{msc}}(\rho_{ij})] \sim -\frac{2}{3}\log_2 N+ \mathrm{const},
\end{aligned}
\end{equation}
thus different from coherence and ASC, the scaling exponents for
the two measures of MSC are different. These scaling behaviors have
also been verified numerically. As shown in the bottommost
panel of Fig. \ref{fig:2}, with the increase of the system size
$N$, the slopes of $\log_2 [C_{l_1}^{\mathrm{msc}}(\rho_{ij})]$
and $\log_2 [C_{r}^{\mathrm{msc}}(\rho_{ij})]$ approach $-1/3$
and $-2/3$, respectively.

\begin{figure}
\centering
\resizebox{0.41 \textwidth}{!}{%
\includegraphics{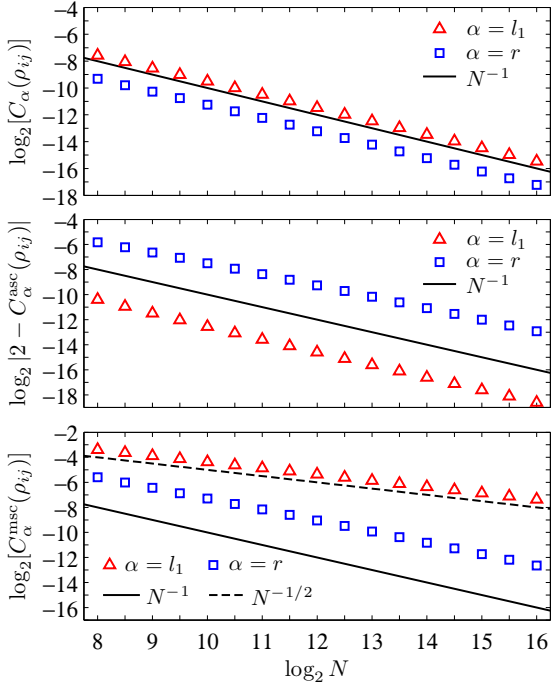}}
\caption{Scaling behaviors of $C_{\alpha}(\rho_{ij})$,
$C_{\alpha}^{\mathrm{asc}}(\rho_{ij})$, and
$C_{\alpha}^{\mathrm{msc}}(\rho_{ij})$ ($\alpha=l_1$ or $r$)
in the symmetric phase with $h=1.1$ and $\gamma=0.5$. The slopes of the
solid and dashed lines are $-1$ and $-1/2$, respectively.} \label{fig:3}
\end{figure}

Next, we consider scaling behaviors of the coherence, ASC,
and MSC in the symmetric phase with $h>1$. By defining
$\Xi=(h-1)(h-\gamma)$, $\gamma_{\pm}= 1\pm \gamma$, and
neglecting the exponentially small terms in $1/N$, the
CUT results of $\langle S_z \rangle$ and
$\langle S_\upsilon^2\rangle$ ($\upsilon=x,y,z$) can be
written as \cite{CUT1}
\begin{equation}\label{eq4-16}
\begin{aligned}
 & \frac{2\langle S_z\rangle}{N} \sim 1+\frac{b_z}{N}, ~
   \frac{4\langle S_x^2\rangle}{N^2} \sim \frac{b_{xx}}{N}, \\
 & \frac{4\langle S_y^2\rangle}{N^2} \sim \frac{b_{yy}}{N}, ~
   \frac{4\langle S_z^2 \rangle}{N^2} \sim 1+\frac{b_{zz}}{N},
\end{aligned}
\end{equation}
with $b_z=1+(\gamma_{+}-2h)/2\Xi^{1/2}$, $b_{xx}=(h-\gamma)/\Xi^{1/2}$,
$b_{yy}=1/b_{xx}$, and $b_{zz}=2b_z$. Then, by combining Eq.
\eqref{eq4-16} with Eqs. \eqref{eq3-3}--\eqref{eq3-6} and after some
algebra, one can obtain
\begin{equation}\label{eq4-17}
\begin{aligned}
 & C_{l_1}(\rho_{ij}) \sim \frac{2b_0-b_z}{N},~
   C_{r}(\rho_{ij}) \sim -\frac{b_z}{N},\\
 & C_{l_1}^\mathrm{asc}(\rho_{ij}) \sim 2 +\frac{2(b_0+b_z)}{N},~
   C_{r}^\mathrm{asc}(\rho_{ij}) \sim 2+\frac{b_z}{N\ln 2},\\
 & C_{l_1}^\mathrm{msc}(\rho_{ij}) \sim \frac{2b_0-b_z}{\sqrt{-2b_z} N^{1/2}},~
   C_{r}^\mathrm{msc}(\rho_{ij}) \sim -\frac{(2b_0-b_z)^2}{8 N b_{z}\ln 2},
\end{aligned}
\end{equation}
where we have defined $b_0=b_{xx}-b_{yy}$ for the conciseness of Eq.
\eqref{eq4-17}. As $b_{xx}>1$, $b_z<0$, and $b_0+b_z>0$, one can find that
\begin{equation}\label{eq4-18}
\begin{aligned}
 & \log_2 [C_{\alpha}(\rho_{ij})]\sim -\log_2 N +\mathrm{const}, \\
 & \log_2 [C_{l_1}^\mathrm{asc}(\rho_{ij})-2] \sim -\log_2 N +\mathrm{const}, \\
 & \log_2 [2-C_{r}^\mathrm{asc}(\rho_{ij})] \sim -\log_2 N +\mathrm{const}, \\
 & \log_2 [C_{l_1}^\mathrm{msc}(\rho_{ij})] \sim -\frac{1}{2}\log_2 N +\mathrm{const}, \\
 & \log_2 [C_{r}^\mathrm{msc}(\rho_{ij})] \sim -\log_2 N +\mathrm{const}.
\end{aligned}
\end{equation}
Thus the scaling exponents for the two measures of coherence and ASC
as well as that for the relative entropy of MSC are all $-1$, whereas
it is $-1/2$ for the $l_1$ norm of MSC, see also Fig. \ref{fig:3}.
This is different from those scaling exponents at the critical point
$h=1$. It indicates that with the increase of the system size $N$,
$C_{\alpha}(\rho_{ij})$ and $C_{\alpha}^\mathrm{msc}(\rho_{ij})$ in
the region of $h>1$ decrease faster than those at $h=1$, while
$C_{r}^\mathrm{asc}(\rho_{ij})$ increases faster than that at $h=1$.
Moreover, $C_{l_1}^\mathrm{asc}(\rho_{ij})$ decreases with the
increase of $N$ and approaches 2 in the thermodynamic limit, which
is opposite to $C_{l_1}^\mathrm{asc}(\rho_{ij})$ at the critical point.

\begin{figure}
\centering
\resizebox{0.41 \textwidth}{!}{%
\includegraphics{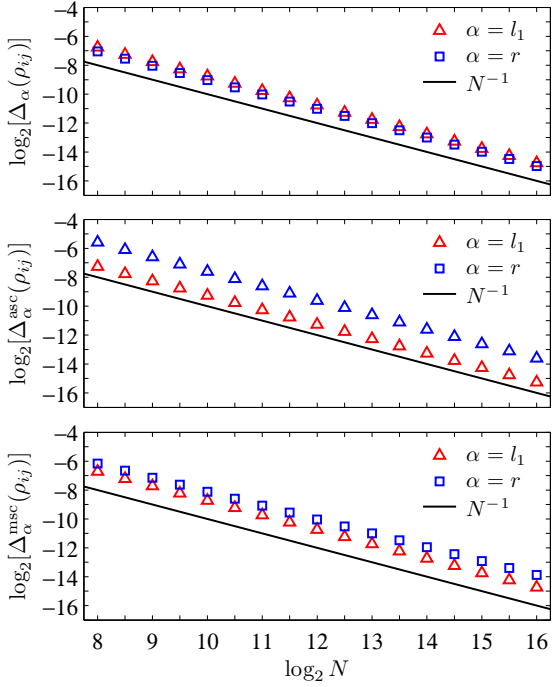}}
\caption{Scaling behaviors of $C_{\alpha}(\rho_{ij})$,
$C_{\alpha}^{\mathrm{asc}}(\rho_{ij})$, and
$C_{\alpha}^{\mathrm{msc}}(\rho_{ij})$ ($\alpha=l_1$ or $r$) in the
broken phase with $h=0.9$ and $\gamma=0.5$, where we have denoted by
$\Delta_{\alpha}(\rho_{ij})= C_{\alpha}(\rho_{ij}^\mathrm{T})- C_{\alpha}(\rho_{ij})$,
$\Delta_{\alpha}^\mathrm{asc}(\rho_{ij})= C_{\alpha}^\mathrm{asc}(\rho_{ij})- C_{\alpha}^\mathrm{asc}(\rho_{ij}^\mathrm{T})$,
and $\Delta_{\alpha}^\mathrm{msc}(\rho_{ij})= C_{\alpha}^\mathrm{msc}(\rho_{ij}^\mathrm{T})- C_{\alpha}^\mathrm{msc}(\rho_{ij})$
in this plot, and the slopes of the solid lines are $-1$.} \label{fig:4}
\end{figure}

Finally, by defining $\Lambda=1-h^2$ and neglecting those exponentially
small terms in $1/N$, the CUT results of $\langle S_z\rangle$ and
$\langle S_\upsilon^2\rangle$ ($\upsilon=x,y,z$) in the broken phase
can be obtained as \cite{CUT1}
\begin{equation}\label{eq4-19}
\begin{aligned}
 & \frac{2\langle S_z\rangle}{N} \sim h+\frac{c_z}{N},~
   \frac{4\langle S_x^2\rangle}{N^2} \sim \Lambda+\frac{c_{xx}}{N}, \\
 & \frac{4\langle S_y^2\rangle}{N^2} \sim \frac{c_{yy}}{N}, ~
   \frac{4\langle S_z^2 \rangle}{N^2} \sim h^2+\frac{c_{zz}}{N},
\end{aligned}
\end{equation}
with $c_z=h(\gamma_{-}/\Lambda)^{1/2}$, $c_{xx}=2+(\gamma h^2+\gamma-2)/
(\Lambda\gamma_{-})^{1/2}$, $c_{yy}=h/c_z$, and $c_{zz}= 2hc_z+
(\Lambda\gamma_{-})^{1/2}$. By combining Eq. \eqref{eq4-19} with Eqs.
\eqref{eq3-3}--\eqref{eq3-6} and after some algebra, one can obtain
\begin{equation}\label{eq4-20}
\begin{aligned}
 & C_{\alpha}(\rho_{ij}) \sim C_{\alpha}(\rho_{ij}^\mathrm{T}) +\frac{a_\alpha}{N},~
   C_{\alpha}^\mathrm{asc}(\rho_{ij}) \sim C_{\alpha}^\mathrm{asc}(\rho_{ij}^\mathrm{T}) +\frac{b_\alpha}{N}, \\
 & C_{\alpha}^\mathrm{msc}(\rho_{ij}) \sim C_{\alpha}^\mathrm{msc}(\rho_{ij}^\mathrm{T}) +\frac{d_\alpha}{N},
\end{aligned}
\end{equation}
where $a_\alpha$, $b_\alpha$, and $d_\alpha$ ($\alpha=l_1$ or
$r$) are complicated polynomials of $\gamma$ and $h$ given by
\begin{equation}\label{eq4-21}
\begin{aligned}
 & a_{l_1}=\frac{c_0-c_{zz}}{2},~ b_{l_1}=\frac{\kappa_2+3c_z+c_0}{2},~d_{l_1}=\frac{c_0-c_{zz}}{2\sqrt{\Lambda}},~\\
 & a_{r}= \frac{\kappa_1}{4}\left[\frac{1}{\ln 2}+\log_2 \left(\frac{1+h^2}{2}\right)\right]-\frac{c_{zz}}{2}\left(1+\frac{1}{\ln 2}\right)\\
 & ~~~~~~~~~+\frac{c_{zz}+2c_z}{4}\log_2\left(\frac{h_+^2}{4}\right)+\frac{c_{zz}-2c_z}{4}\log_2\left(\frac{h_-^2}{4}\right), \\
 & b_{r}=\frac{\kappa_2}{2}\log_2\left(\frac{\kappa_{+}}{\kappa_{-}}\right),~
   d_{r}=\kappa_3\log_2\left(\frac{h_{-}}{h_{+}}\right)+\frac{2\kappa_3 h+a_{l_1}}{2\ln 2},
\end{aligned}
\end{equation}
where $c_0= c_{xx}-c_{yy}$, $\kappa_{\pm}= 1\pm (\Lambda^2+h^2)^{1/2}$,
$\kappa_1= c_{zz}+(\Lambda c_0+4h c_z)/(1+h^2)$,
$\kappa_2= (h c_z+ \Lambda a_{l_1})/(\Lambda^2+h^2)^{1/2}$, and
$\kappa_3=(2c_z-h c_{zz})/2\Lambda$. Thereby, the following scaling formulas
hold:
\begin{equation}\label{eq4-22}
\begin{aligned}
 & \log_2 [C_{\alpha}(\rho_{ij}^\mathrm{T})-C_{\alpha}(\rho_{ij})] \sim -\log_2 N +\mathrm{const}, \\
 & \log_2 [C_{\alpha}^\mathrm{asc}(\rho_{ij})-C_{\alpha}^\mathrm{asc}(\rho_{ij}^\mathrm{T})] \sim -\log_2 N +\mathrm{const}, \\
 & \log_2 [C_{\alpha}^\mathrm{msc}(\rho_{ij}^\mathrm{T})- C_{\alpha}^\mathrm{msc}(\rho_{ij})] \sim -\log_2 N +\mathrm{const}.
\end{aligned}
\end{equation}
Thus the scaling exponents are always $-1$ in the broken phase,
which are different from those at the QPT point. Such an
observation has been confirmed by the numerical results shown in
Fig. \ref{fig:4}. It can be seen that the prediction of Eq.
\eqref{eq4-22} works quite well. It also indicates that
$C_{\alpha}(\rho_{ij})$ and $C_{\alpha}^\mathrm{msc}(\rho_{ij})$
increase with the increase of the system size $N$ and approach
gradually their thermodynamic limit values of
$C_{\alpha}(\rho_{ij}^\mathrm{T})$ and
$C_{\alpha}^\mathrm{msc}(\rho_{ij}^\mathrm{T})$, respectively,
while $C_{\alpha}^\mathrm{asc}(\rho_{ij})$ decreases with the
increase of the system size $N$ and approaches gradually 
$C_{\alpha}^\mathrm{asc}(\rho_{ij}^\mathrm{T})$.

As the single-spin state $\rho_i$ is concerned, it is
straightforward to obtain from Eqs. \eqref{eq4-6}, \eqref{eq4-16},
and \eqref{eq4-19} that in the symmetric phase, the maximum
attainable coherence scales as
\begin{equation}\label{eq4-n23}
 \log_2 [C_{\alpha}^{\max}(\rho_{i}^\mathrm{T})-C_{\alpha}^{\max}(\rho_{i})] \sim s \log_2 N +\mathrm{const}, \\
\end{equation}
where $s=-2/3$ ($s=-1$) for $h=1$ ($h> 1$), and in the broken phase it
scales as
\begin{equation}\label{eq4-n24}
 \log_2 [C_{\alpha}^{\max}(\rho_{i})- C_{\alpha}^{\max}(\rho_{i}^\mathrm{T})] \sim -\log_2 N +\mathrm{const}.
\end{equation}
Thus the scaling exponents for the maximum coherence of the
single-spin state are completely the same as those for
$\langle S_z \rangle/N$. This is understandable as they
are uniquely determined by the magnetization intensity
$\langle S_z\rangle$.

\begin{figure}
\centering
\resizebox{0.41 \textwidth}{!}{%
\includegraphics{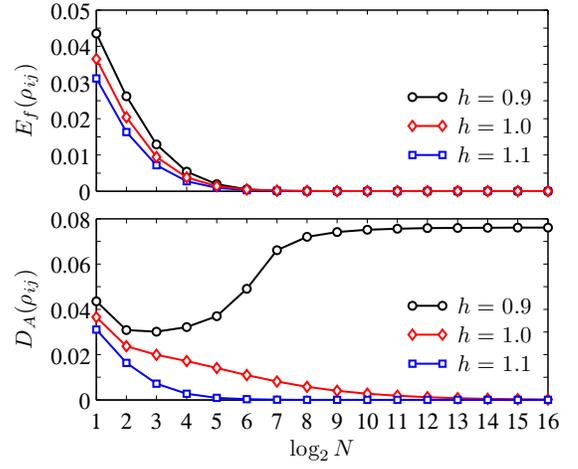}}
\caption{Entanglement $E_f(\rho_{ij})$ and quantum discord $D_A(\rho_{ij})$
versus $N$ with fixed $\gamma=0.5$ and different $h$.} \label{fig:5}
\end{figure}

Before ending this section, we provide a brief comparison
of the performance of the coherence based indicators with the
other quantum correlations, including entanglement of formation
$E_f(\rho_{ij})$ \cite{EoF1,EoF2} and quantum discord
$D_A(\rho_{ij})$ \cite{QD1,QD2} of $\rho_{ij}$. As is shown in
Fig. \ref{fig:5}, $E_f(\rho_{ij})$ decays rapidly with the
increase of $N$ and becomes extremely small after $N\gtrsim 2^6$,
while for $D_A(\rho_{ij})$ it also decreases rapidly with the
increase of $N$ in the symmetric phase and approaches a very
weak value in the broken phase. This may reduce their effectiveness
in signaling the criticality of the LMG model as the accurate
measurement of such weak quantities is a highly demanding task in
experiments. But as what we have shown above, the coherence based
indicators do not suffer from this problem, and this might be
considered an advantage of them over entanglement and discord.
Moreover, quantum coherence and ASC are analytically solvable for
any state \cite{coher,asc1}, the coherence can be estimated with
elaborately designed experiments \cite{exp1,exp2}, and the
measurement of ASC is also comparably simpler than that of
entanglement and discord as it does not require a two-spin state
tomography. Finally, the coherence based indicators are not so
sensitive to the increasing system size $N$ (they are finite even
for infinite $N$) and the choice of two spins in the system. This
property is also valuable as it removes the restriction on the
necessity of a careful choice of two spins. Thus, the coherence
based measures might serve as complementary indicators of
quantum criticality in many-body systems.

\section{Summary and discussion} \label{sec:5}
To summarize, we have investigated the coherence, ASC, and MSC of
the LMG model which undergoes a second-order QPT with the variation
of the transverse magnetic field. The motivation for considering
this problem lies in that coherence reflects the origin of quantumness
and underlies various forms of quantum correlations, thus QPTs which
correspond to dramatic changes of the ground state of a system, may
be tied to the critical changes of the coherence based quantifiers.
Moreover, the properties of the high-dimensional systems are rich
compared to those one-dimensional ones in general, while the
coherence based indicators are insensitive to the system size, thus
they may be capable of capturing some key ingredients of the quantum
criticality, especially for those in the regions with vanishing
entanglement and discord. Thereby, it is worthy to present a
thorough analysis of them.

We have obtained the analytical solutions of the coherence, ASC,
and MSC for the thermodynamic limit and the isotropic cases of the
LMG model, and analyzed in detail their properties by using the
exact diagonalization method for the general anisotropic case. The
results showed that they exhibit distinct behaviors in the symmetric
and broken phases. These distinct behaviors could be identified as
reliable indicators of quantum criticality in the LMG model.
They also confirm that these coherence based indicators work
well in signaling quantum criticality not only for the one-dimensional
spin systems \cite{Karpat,Qin,Leisg,Liyc,Ywl,spin1} but also for the
high-dimensional systems.

We have also obtained finite-size scaling exponents for the
coherence, ASC, and MSC by using the CUT method and confirmed them
numerically with the system size up to $N=2^{16}$. It is found that
their dependence on $N$ is also phase dependent. Specifically, the
two measures of coherence and MSC as well as the relative entropy
of ASC show opposite dependence on $N$ in the broken and symmetric
phases, whereas the $l_1$ norm of ASC shows opposite dependence on $N$
at and beyond the QPT point. Moreover, the differences between
their scaling exponents can be comprehend as follows. First, for the
single-spin state, the maximum coherence is solely determined
by $\langle S_z \rangle$, so the scaling exponent is in accord with
that of $2\langle S_z\rangle/N$. Second, the coherence, ASC, and MSC
for the two-spin state are determined by $\langle S_z\rangle$ and
$\langle S_\upsilon^2\rangle$ ($\upsilon=x,y,z$). In the broken
phase and the symmetric phase with $h>1$, the scaling exponents for
$2\langle S_z \rangle/N$ and $4\langle S_\upsilon^2\rangle/N^2$ are
always $-1$, while at the critical point $h=1$, the scaling exponents
for $2\langle S_z \rangle/N$ and $4\langle S_{x,z}^2 \rangle/N^2$ are
$-2/3$ and that for $4\langle S_y^2 \rangle/N^2$ is $-4/3$ (this term
can be neglected as it is much smaller than the other terms). These
lead to the same scaling exponents for the two measures of coherence
and ASC as well as the relative entropy of MSC as those of
$2\langle S_z\rangle/N$ and $4\langle S_{x,z}^2 \rangle/N^2$. As for
the $l_1$ norm of MSC, due to the square root in Eq. \eqref{eq3-5},
its scaling exponent in the symmetric phase equals half of those
for $2\langle S_z \rangle/N$ and $4\langle S_{x,z}^2 \rangle/N^2$,
while in the broken phase, $1-(v_1-v_2)^2\sim \Lambda$, thus its
scaling exponent is still the same as those for
$2\langle S_z \rangle/N$ and $4\langle S_\upsilon^2 \rangle/N^2$.
Physically, the $l_1$ norm of ASC shows a different dependence on $N$
from the other coherence based indicators and the $l_1$ norm of MSC
does not recover the underlying critical exponents of
$2\langle S_z\rangle/N$ and $4\langle S_\upsilon^2 \rangle/N^2$ in
the symmetric phase having their roots in the $l_1$ norm of coherence
given in Eq. \eqref{eq2-1} as it neglects the diagonal elements of
$\rho_{ij}$. In this sense, we argue that the relative entropy of
coherence based indicators is more beneficial for studying the
quantum criticality.

While the information uncovered by coherence and steered coherence
may provide alternative perspective for understanding the quantum
criticality in the LMG model, one might also concern the experimental
verification of the connections between QPTs and coherence based
indicators established here. For this purpose, one may resort to
the progress on simulating the LMG model in trapped ions
\cite{lmgsm1,lmgsm2}, nitrogen-vacancy center ensembles
\cite{lmgsm3}, and superconducting qubits \cite{lmgsm4}. In
particular, the dynamical phase transition in the LMG model has been
experimentally demonstrated in a quantum simulator with all-to-all
connected superconducting qubits \cite{lmgsm4}. Hence, it is
possible to expect verification of these connections in 
future experiments with similar platforms. Moreover, the
coherence based indicators require no prior knowledge of the order
parameters; thus it is also of great interest to use them to study
the exotic quantum phases in many-body systems such as the
topological ordered phase without any local order parameter
\cite{top1,top2,top3}. Additionally, the coherence based indicators
are finite for almost all states (e.g., the ASC vanishes only for
the maximally mixed state), even when discord is absent \cite{Hu},
so it is also appealing to use them to study the quantum criticality
in the parameter regions without entanglement and discord, e.g., the
factorization phenomenon which corresponds to the occurrence of a
fully factorized ground state at certain critical driving system
parameters \cite{Karpat,Ywl,fac1,fac2,fac3}.

\section*{ACKNOWLEDGMENTS}
This work was supported by the National Natural Science Foundation
of China (Grant Nos. 11675129 and 11934018), the Strategic Priority
Research Program of Chinese Academy of Sciences (Grant No. XDB28000000),
and Beijing Natural Science Foundation (Grant No. Z200009).

\begin{appendix}
\section{Derivation of the MSC} \label{sec:A}
\setcounter{equation}{0}
\renewcommand{\theequation}{A\arabic{equation}}

As the reduced density operator $\rho_j= \tr_i \rho_{ij}$ is
nondegenerate for $h\neq 0$, one only needs to take the
maximization over the set of projective measurements $M=(\iden+
\vec{m}\cdot\vec{\sigma})/2$ (see \cite{msc} for an explanation),
where $\vec{\sigma}=(\sigma_x,\sigma_y,\sigma_z)$ is a vector
composed of the Pauli operators and $\vec{m}$ is a unit vector in
$\mathbb{R}^3$ with the polar and azimuthal angles denoted by
$\vartheta$ and $\varphi$, respectively.

For $C_{l_1}^{\mathrm{asc}}(\rho_{ij})$, from Ref. \cite{msc} one
can obtain
\begin{equation}\label{eqa-1}
\begin{aligned}
 C_{l_1}^{\mathrm{msc}}(\rho_{ij})= \max_{\{\vartheta,\varphi\}}
                                    \frac{\sqrt{T_{11}^2\cos^2\varphi+ T_{22}^2\sin^2\varphi}\sin\vartheta}
                                    {1+(v_1-v_2)\cos\vartheta},
\end{aligned}
\end{equation}
where $T_{mn}= \tr(\rho_{ij}\sigma_m\otimes\sigma_n)$. It is direct
to obtain that $T_{11}= 2(y+ u)$ and $T_{22}= 2(y- u)$ (hence
$T_{11}^2> T_{22}^2$), and the optimal azimuthal angle is given by
$\varphi_0= 0$ or $\pi$. In addition, one can obtain directly the
optimal polar angle as $\vartheta_0= \arccos(v_2-v_1)$. All these
yield $C_{l_1}^{\mathrm{msc}}(\rho_{ij})$ in Eq. \eqref{eq3-5}.

For $C_{r}^{\mathrm{msc}}(\rho_{ij})$, we derive its
expression as follows. First, the eigenbasis of
$\rho_j=\tr_i\rho_{ij}$ is given by $\{\mid\uparrow\rangle, \mid\downarrow\rangle\}$.
Within this basis, the postmeasurement state of spin $j$ can be
obtained as
\begin{eqnarray}\label{eqa-2}
 \rho_{j|M}= \frac{1}{p_M}
  \begin{pmatrix}
   v_1\cos^2\frac{\vartheta}{2}+y\sin^2\frac{\vartheta}{2}   & \frac{\sin\vartheta}{2}(u e^{i\varphi}+y e^{-i\varphi}) \vspace{0.5em} \\
   \frac{\sin\vartheta}{2} (u e^{-i\varphi}+y e^{i\varphi})  & v_2\sin^2\frac{\vartheta}{2}+y\cos^2\frac{\vartheta}{2}  \\
  \end{pmatrix},
\end{eqnarray}
where $p_M=y+v_2+ (v_1-v_2)\cos^2(\vartheta/2)$. Hence, according to
its definition \cite{msc}, the MSC is given by
\begin{equation}\label{eqa-3}
 C_{r}^{\mathrm{asc}}(\rho_{ij})= \max_{\{\vartheta,\varphi\}}
                                            \left\{ S[(\rho_{j|M})_\mathrm{diag}]-S(\rho_{j|M})\right\}.
\end{equation}
Then after some algebra, one can obtain that the optimal polar angle
and azimuthal angle related to $M$ are still given by $\vartheta_0=
\arccos(v_2-v_1)$ and $\varphi_0= \{0,\pi\}$, respectively. By
inserting them into Eq. \eqref{eqa-2}, one can obtain the optimized
$\rho_{j|M}$ (we denote by $r_{mn}$ its element that lies in the $m$th
row and $n$th column). Hence, we obtained
$C_{r}^{\mathrm{asc}}(\rho_{ij})$ given in Eq. \eqref{eq3-5}.

\end{appendix}

\newcommand{\PRL}{Phys. Rev. Lett. }
\newcommand{\RMP}{Rev. Mod. Phys. }
\newcommand{\PRA}{Phys. Rev. A }
\newcommand{\PRB}{Phys. Rev. B }
\newcommand{\PRD}{Phys. Rev. D }
\newcommand{\PRE}{Phys. Rev. E }
\newcommand{\PRX}{Phys. Rev. X }
\newcommand{\NJP}{New J. Phys. }
\newcommand{\JPA}{J. Phys. A }
\newcommand{\JPB}{J. Phys. B }
\newcommand{\OC}{Opt. Commun.}
\newcommand{\PLA}{Phys. Lett. A }
\newcommand{\EPJB}{Eur. Phys. J. B }
\newcommand{\EPJD}{Eur. Phys. J. D }
\newcommand{\NP}{Nat. Phys. }
\newcommand{\NC}{Nat. Commun. }
\newcommand{\EPL}{Europhys. Lett. }
\newcommand{\AdP}{Ann. Phys. (Berlin) }
\newcommand{\AoP}{Ann. Phys. (N.Y.) }
\newcommand{\QIC}{Quantum Inf. Comput. }
\newcommand{\QIP}{Quantum Inf. Process. }
\newcommand{\CPB}{Chin. Phys. B }
\newcommand{\IJTP}{Int. J. Theor. Phys. }
\newcommand{\IJQI}{Int. J. Quantum Inf. }
\newcommand{\IJMPB}{Int. J. Mod. Phys. B }
\newcommand{\PR}{Phys. Rep. }
\newcommand{\SR}{Sci. Rep. }
\newcommand{\LPL}{Laser Phys. Lett. }
\newcommand{\SCG}{Sci. China Ser. G }
\newcommand{\JMP}{J. Math. Phys. }
\newcommand{\RPP}{Rep. Prog. Phys. }
\newcommand{\PA}{Physica A }


\end{document}